\documentclass[aps,prd,preprint,amsmath,amssymb,
nofootinbib,eqsecnum,showpacs,preprint,tightenlines]{revtex4}
\usepackage{epsfig}
\usepackage{rotating}
\usepackage{bm}
\newcommand{\be}{\begin{equation}}
\newcommand{\ee}{\end{equation}}
\newcommand{\bea}{\begin{eqnarray}}
\newcommand{\eea}{\end{eqnarray}}
\newcommand{\nn}{\nonumber}
\def\tEo{\tilde{{\mathcal E}}^{(1)}}
\def\tEt{\tilde{{\mathcal E}}^{(2)}}

\begin{document}
\title{Local and Global Casimir Energies for a Semitransparent 
Cylindrical Shell}
\author{In\'es Cavero-Pel\'aez}
\email{cavero@nhn.ou.edu}

\author{Kimball A. Milton}
\email{milton@nhn.ou.edu}
\homepage{www.nhn.ou.edu/

\affiliation{Oklahoma Center for High-Energy Physics and Homer L. Dodge 
Department of Physics and Astronomy, University of Oklahoma,
Norman, OK 73019-2061}

\author{Klaus Kirsten}
\email{Klaus_Kirsten@baylor.edu}
\affiliation{Department of Mathematics, Baylor University, Waco, TX 76798}

\date{\today}

\pacs{03.70.+k, 11.10.Gh, 03.65.Sq, 11.30.Cp}

\begin{abstract}
The local Casimir energy density and the global Casimir energy
for a massless scalar field
associated with a $\lambda\delta$-function potential in a $3+1$
dimensional circular cylindrical geometry are considered.  The global
energy is examined for both weak and strong coupling, the latter
being the well-studied Dirichlet cylinder case.  For weak-coupling,
through $\mathcal{O}(\lambda^2)$, 
the total energy is shown to vanish by both analytic
and numerical arguments, based both on Green's-function and zeta-function
techniques.  Divergences occurring in the calculation are
shown to be absorbable by renormalization of physical parameters of the
model.  The global energy may be obtained by integrating 
the local energy density
only when the latter is supplemented by an energy term residing precisely
on the surface of the cylinder.  The latter is identified as the integrated 
local energy density of the cylindrical shell when the latter is physically
expanded to have finite thickness.  Inside and outside the $\delta$-function
shell, the local energy density diverges as the surface of the
shell is approached; the divergence
is weakest when the conformal stress tensor is used to define the energy
density.  A real global divergence first occurs in $\mathcal{O}(\lambda^3)$, as
anticipated, but the proof is supplied here for the first time; this
divergence is entirely associated with the surface energy, and does {\em not\/}
reflect divergences in the local energy density as the surface is approached.

\end{abstract}

\maketitle
\section{Introduction}
The subject of self-energies due to quantum fluctuations of fields constrained
by physical boundaries, such as the Casimir energy due to a perfectly conducting
spherical shell \cite{Boyer:1968uf}, has been controversial nearly
from the outset \cite{Deutsch:1978sc,Candelas:1981qw}.  Partly this is because
such a self-energy is apparently not well-defined, for it is not accessible
by deforming part of the surface.  More fundamentally, it is because there
are strong divergences in the local energy density as the (idealized) boundary
is approached, which would seem to rule out the existence of a finite total
energy for the system.  How are such divergences to be squared with the
mathematical proof that the electromagnetic Casimir energy of a region
with a closed, smooth, perfectly conducting boundary is finite \cite{graf}?

These issues have been brought to the forefront by a series of papers by
Graham et al.~\cite{Graham:2003ib}. Essentially,
they assert that it is impossible to ascribe any physical meaning to the
self-Casimir energy of a single object, such as Boyer's sphere 
\cite{Boyer:1968uf}. However, the divergence issues they raise are hardly
new \cite{Milton:1979yx}; for example, for a massless scalar particle in
the presence of a spherical $\delta$-shell potential, governed by the
Lagrangian
\be
\mathcal{L}=-\frac12\partial_\mu\phi\partial^\mu\phi-\frac12
\frac{\lambda}{a^2}\delta(r-a)\phi^2,\label{deltapot}
\ee
a divergence occurs in third-order in $\lambda$  \cite{Bordag:1998vs}, 
and possible ways of dealing with it
have been suggested \cite{Bordag:2004rx,Scandurra:1998xa}.

For a different viewpoint see the considerations of Barton \cite{barton}. He
argues, that for any {\it single connected body} the physics is
dominated by the divergent components of the self-energy. The pure
Casimir terms, which in the examples studied in Ref.~\cite{barton} 
are the convergent components when the no-cutoff limit is taken, 
are shown to be much smaller such that they will never become observable.

In a recent paper \cite{Cavero-Pelaez:15kq} the model described by
(\ref{deltapot}) was considered in some detail.  In fact, the model
examined was somewhat more general, in that the $\delta$ function was replaced
by a step-function potential of width $\delta$ and height $h$.  As $\delta
\to0$ and $h\to\infty$ with $h\delta$ held fixed at unity, a $\delta$-function
potential is recovered.  For such a potential we calculated the energy
density inside and outside the region of the potential, from which we
could calculate the total energy as a function of $\lambda$, as well as
the local energy density.  As long as the potential is finite, the local
energy density may be integrated to yield the total energy.  In the singular
$\delta$-function limit, however, the energy within the shell becomes a
localized surface energy which must be added to the integrated local energy
of the regions inside ($r<a$) and outside ($r>a$) the sphere, which has
precisely the anticipated form 
\cite{Dowker:1978md,Kennedy:1979ar,saharian,fulling}.  For weak coupling with
the $\delta$-shell potential,
the total Casimir energy is finite in $\mathcal{O}(\lambda^2)$, 
but divergent in
third order, which divergence precisely corresponds to the divergence of
the surface energy in that order.  Thus, it is plausible that such a
divergence should be absorbed in a renormalization of the surface energy.

In this paper, we turn to the corresponding cylindrical case.  The situation in
many ways is similar.  However, there are curiosities associated with
the cylindrical geometry that make this new analysis intriguing.
First, the Casimir self-stress on a perfectly conducting circular
cylinder was found \cite{DeRaad:1981hb} to be
attractive, and of somewhat smaller magnitude, compared to the repulsive
stress found by Boyer for a perfectly conducting sphere \cite{Boyer:1968uf}.
It was found that a dilute dielectric cylinder had vanishing van der Waals
energy \cite{romeo, nesterenko}, as did a dilute dielectric-diamagnetic
cylinder (with the speed of light the same on the inside and the outside)
\cite{nesterenko,Nesterenko:1999zg,Klich:1999ip}, 
which seemed to imply the vanishing of the Casimir energy for a dielectric 
cylinder of permittivity $\varepsilon$ in order $(\varepsilon-1)^2$, as was
only recently verified \cite{Cavero-Pelaez:2004xp,Romeo:2005qk}.  If
a perfectly conducting cylinder is slightly deformed by giving its cross 
section
a slight eccentricity $\delta e$, the change in the Casimir energy vanishes
in order $\delta e^2$ \cite{kitsonromeo}.  None of these vanishings occur for
a sphere \cite{Milton:1997ky,Brevik:1998zs,barton99,hoye00,
Klich:1999df,Kitson:2005kk}, 
so they all reflect the flatness of the cylindrical geometry.  In this paper
we establish another example of the second-order vanishing effect for the
cylinder, that is, that the $\mathcal{O}(\lambda^2)$ 
term in the Casimir energy for
the semitransparent cylinder is zero.  A summary of the facts comparing
results for sphere and cylinder, together with the first reference for
each result, is given in Table \ref{tab1}.

\begin{table}
\begin{tabular}{cccl}
Type&$E_{\rm Sphere}a$&$\mathcal{E}_{\rm Cylinder}a^2$&References\\
\hline\\
EM&$+0.04618$&$-0.01356$&\cite{Boyer:1968uf} \cite{DeRaad:1981hb}\\
D&$+0.002817$&$+0.0006148$&\cite{Bender:1994zr}\cite{gosrom}\\
$(\varepsilon-1)^2$&$+0.004767=\frac{23}{1536\pi}$&$0$
&\cite{Brevik:1998zs}\cite{Cavero-Pelaez:2004xp}\\
$\xi^2$&$+0.04974=\frac5{32\pi}$&$0$&\cite{Klich:1999df}\cite{nesterenko}\\
$\delta e^2$&$\pm0.0009$&$0$&\cite{Kitson:2005kk}\cite{kitsonromeo}\\
$\lambda^2$&$+0.009947=\frac1{32\pi}$&$0$&\cite{milton03}\\
\hline\\
\end{tabular}
\label{tab1}
\caption{Casimir energy ($E$) for a sphere and Casimir energy per unit
length ($\mathcal{E}$) for a cylinder, both of radius $a$.  
Here the different boundary
conditions are perfectly conducting for electromagnetic fields (EM),
Dirichlet for scalar fields (D), dilute dielectric for electromagnetic
fields [coefficient of $(\varepsilon-1)^2$], dilute dielectric for 
electromagnetic fields
with media having the same speed of light (coefficient of $\xi^2
=[(\varepsilon-1)/(\varepsilon+1)]^2$), 
for perfectly conducting surface with eccentricity $\delta e$
(coefficient of $\delta e^2$), and weak coupling
for scalar field with $\delta$-function boundary given by (\ref{deltapot}),
(coefficient of $\lambda^2$).  The references given are, to the authors' 
knowledge, the
first paper in which the results in the various cases were found.}
\end{table}

The outline of this paper is as follows.  In the next section we
will derive the Green's function for the semitransparent cylinder.
In Sec.~\ref{sec:pressure} we will then compute the Casimir pressure on
the cylinder, and thereby infer the Casimir energy.  In Sec.~\ref{sec:energy}
we will rederive that energy directly.  The weak-coupling evaluation of the
Casimir energy will be the subject of Sec.~\ref{sec:weak}.  We will show that
the Casimir energy through order $\lambda^2$ vanishes both by analytic and
numerical arguments, with an explicit isolation of the divergent term which
may be unambiguously removed.  An independent derivation of these results
using zeta function techniques is given in Sec.~\ref{sec:zf}.  
The proof that for a cylinder of arbitrary cross section the divergence in the
Casimir energy occurs in $\mathcal{O}(\lambda^3)$ is supplied in Sec.~\ref{hk}.
As we have noted in Table \ref{tab1},
the strong-coupling (Dirichlet) result was earlier derived by Gosdzinsky and
Romeo \cite{gosrom}; we reproduce their result in Sec.~\ref{sec:sc}.
Again the explicit divergent terms that must be removed by renormalization
are identified.  We turn to an examination of the local energy density in
Sec.~\ref{sec:local}.  We show that the integrated local energy density
differs from the total energy by a surface term that resides precisely on
the cylindrical surface.  We also examine the surface divergences that
appear in the energy density as one approaches the surface; the leading
divergence, which is independent of the shape of the surface, may be eliminated
by choosing the conformal stress tensor.  Then one is left with surface
divergences,
in either strong or weak coupling, that are exactly one-half that for a sphere,
reflecting the vanishing of one of the principal curvatures for a cylinder.
In Sec.~\ref{sec:thick}, we examine what happens when the 
$\delta$-function potential is thickened to a cylindrical annulus of thickness
$\delta$ and height $h$.  We show that then the surface energy is resolved
as the integrated local energy density of the field within the confines of
the annulus.  The divergence in the total energy in $\mathcal{O}(\lambda^3)$
for a $\delta$-function shell is exactly that due to the surface energy alone.
 Finally, we offer a perspective of the situation in the 
Conclusion.  Appendix A offers further details on the divergences
occuring in the global theory, particularly in $\mathcal{O}(\lambda)$, while
Appendix B elaborates some further aspects of the surface
energy.

\section{Green's Function}
\label{sec:gf}
We consider a massless scalar field $\phi$ in a $\delta$-cylinder background,
\be
\mathcal{L}_{\rm int}=-\frac\lambda{2a}\delta(r-a)\phi^2,
\ee
$a$ being the radius of the ``semitransparent'' cylinder.  We recall that the
massive case was earlier considered by Scandurra \cite{scandurra}.  Note that
with this definition, $\lambda$ is dimensionless.
The time-Fourier transform of the Green's function,
\be
G(x,x')=\int\frac{d\omega}{2\pi}e^{-i\omega(t-t')}\mathcal{G}(\mathbf{r,r'}),
\ee
satisfies
\be
\left[-\nabla^2-\omega^2+\frac\lambda{a}\delta(r-a)\right]\mathcal{G}
(\mathbf{r,r'})=\delta(\mathbf{r-r'}).
\ee
Adopting cylindrical coordinates, we write
\be
\mathcal{G}(\mathbf{r,r'})=\int\frac{dk}{2\pi}e^{ik(z-z')}
\sum_{m=-\infty}^\infty
\frac1{2\pi}e^{im(\varphi-\varphi')}g_m(r,r';k),
\ee
where the reduced Green's function satisfies
\be
\left[-\frac1r\frac{d}{dr}r\frac{d}{dr}+\kappa^2+\frac{m^2}{r^2}
+\frac\lambda{a}
\delta(r-a)\right]g_m(r,r';k)=\frac1r\delta(r-r'),\label{rgfeqn}
\ee
where $\kappa^2=k^2-\omega^2$.  Let us immediately make a Euclidean rotation,
\be
\omega\to i\zeta,
\ee
where $\zeta$ is real, so $\kappa$ is likewise always real.  Apart from
the $\delta$ functions, this is the modified Bessel equation.

Because of the Wronskian satisfied by the modified Bessel functions,
\be
K_m(x)I'_m(x)-K'_m(x)I_m(x)=\frac1x,\label{wronskian}
\ee
we have the general solution to (\ref{rgfeqn}) as long as $r\ne a$ to be
\be
g_m(r,r';k)=I_m(\kappa r_<)K_m(\kappa r_>)+A(r')I_m(\kappa r)+B(r')
K_m(\kappa r),
\ee
where $A$ and $B$ are arbitrary functions of $r'$. 
Now we incorporate the effect of the $\delta$ function at $r=a$
in (\ref{rgfeqn}).
It implies that $g_m$ must be continuous at $r=a$, while it has a discontinuous
derivative,
\be
a\frac{d}{dr}g_m(r,r';k)\bigg|_{r=a-}^{r=a+}=\lambda g_m(a,r';k),
\ee
from which we rather immediately deduce the form of the Green's function
inside and outside the cylinder:
\begin{subequations}
\bea
r,r'<a:\quad g_m(r,r';k)&=&I_m(\kappa r_<)K_m(\kappa r_>)\nonumber\\
&&\quad\mbox{}-\frac{\lambda
K_m^2(\kappa a)}{1+\lambda I_m(\kappa a)K_m(\kappa a)}I_m(\kappa r)
I_m(\kappa r'),\label{gin}
\\
r,r'>a:\quad g_m(r,r';k)&=&I_m(\kappa r_<)K_m(\kappa r_>)\nonumber\\
&&\quad\mbox{}-\frac{\lambda
I_m^2(\kappa a)}{1+\lambda I_m(\kappa a)K_m(\kappa a)}K_m(\kappa r)
K_m(\kappa r').\label{gout}
\eea
\end{subequations}
Notice that in the limit $\lambda\to\infty$ we recover the Dirichlet cylinder
result, that is, that $g_m$ vanishes at $r=a$.

\section{Pressure}
\label{sec:pressure}

The easiest way to calculate the total energy is to compute the pressure on the
cylindrical walls due to the quantum fluctuations in the field.  This may be
computed, at the one-loop level, from the vacuum expectation value of the
stress tensor,
\be
\langle T^{\mu\nu}\rangle
=\left(\partial^\mu\partial^{\prime\nu}-\frac12 g^{\mu\nu}
\partial^\lambda\partial'_\lambda\right)\frac1i G(x,x')\bigg|_{x=x'}-\xi
(\partial^\mu\partial^\nu-g^{\mu\nu}\partial^2)\frac1i G(x,x).\label{st}
\ee
Here we have included the conformal parameter $\xi$, which is equal to 1/6
for the stress tensor that makes conformal invariance manifest.  The conformal 
term does not contribute to the radial-radial component of the stress tensor, 
however, because then only transverse and time derivatives act on $G(x,x)$, 
which depends only on $r$.  The discontinuity of the expectation value of the 
radial-radial   
component of the stress tensor is the pressure of the cylindrical wall:
\bea
P&=&\langle T_{rr}\rangle_{\rm in}-\langle T_{rr}\rangle_{\rm out}\nonumber\\
&=&-\frac1{16\pi^3}\sum_{m=-\infty}^\infty \int_{-\infty}^\infty dk
\int_{-\infty}
^\infty d\zeta\frac{\lambda \kappa^2}{1+\lambda I_m(\kappa a)K_m(\kappa a)}
\nonumber\\
&&\qquad\times
\left[K_m^2(\kappa a)I_m^{\prime 2}(\kappa a)-I_m^2(\kappa a)K_m^{\prime 2}
(\kappa a)\right]\nonumber\\
&=&-\frac1{16\pi^3}\sum_{m=-\infty}^\infty\int_{-\infty}^\infty dk
\int_{-\infty}
^\infty d\zeta\frac\kappa{a}\frac{d}{d\kappa a}\ln\left[1+\lambda I_m(\kappa a)
K_m(\kappa a)\right],
\eea
where we have again used the Wronskian (\ref{wronskian}).  Regarding $ka$ and
$\zeta a$ as the two Cartesian components of a two-dimensional vector, with
magnitude $x\equiv\kappa a=\sqrt{k^2a^2+\zeta^2 a^2}$, we get the stress on the
cylinder per unit length to be
\be
\mathcal{S}=2\pi a P=-\frac1{4\pi a^3}\int_0^\infty dx\,x^2\sum_{m=-\infty}^\infty
\frac{d}{dx}\ln\left[1+\lambda I_m(x)K_m(x) \right],
\ee
which possesses the expected Dirichlet limit as $\lambda\to\infty$.
The corresponding expression for the total Casimir energy per unit length 
follows by integrating
\be
\mathcal{S}=-\frac\partial{\partial a}\mathcal{E},
\ee
that is,
\be
\mathcal{E}=-\frac1{8\pi a^2}\int_0^\infty dx\,x^2\sum_{m=-\infty}^\infty
\frac{d}{dx}\ln\left[1+\lambda I_m(x)K_m(x) \right].\label{energy}
\ee
This expression is of course, completely formal, and will be regulated in
various ways in the following, for example, with an exponential
regulator (in Secs.~\ref{sec:weak}, \ref{sec:sc}, Appendix A), or by using 
zeta-function regularization (in Sec.~\ref{sec:zf}).

\section{Energy}
\label{sec:energy}
Alternatively, we may compute the energy directly from the general
formula \cite{milton04a}
\be
E=\frac1{2i}\int(d\mathbf{r})\int\frac{d\omega}{2\pi}2\omega^2\mathcal{G}
(\mathbf{r,r}).\label{genen}
\ee
To evaluate (\ref{genen}) in this case, we need the indefinite integrals
\begin{subequations}
\bea
\int_0^x dy\,y \,I_m^2(y)&=&
\frac12\left[(x^2+m^2)I_m^2(x)-x^2 I_m^{\prime 2}\right],\label{int1}
\\
\int_x^\infty dy\,y \,K_m^2(y)&=&-\frac12\left[(x^2+m^2)K_m^2(x)-
x^2 K_m^{\prime 2}\right].\label{int2}
\eea
\end{subequations}
When we insert the above construction of the Green's function, and perform
the integrals as indicated over the regions interior and exterior to the
cylinder we obtain
\be
\mathcal{E}=-\frac{a^2}{8\pi^2}\sum_{m=-\infty}^\infty
\int_{-\infty}^\infty d\zeta\int_{-\infty}^\infty dk\,
\zeta^2\frac1x\frac{d}{dx}\ln\left[1+\lambda I_m(x)K_m(x)\right].
\ee
Again we regard the two integrals as over Cartesian coordinates, and replace
the integral measure by
\be
\int_{-\infty}^\infty d\zeta\int_{-\infty}^\infty dk \,\zeta^2=\pi\int_0^\infty
d\kappa\,\kappa^3.\label{angint}
\ee
The result (\ref{energy}) immediately follows. 

\section{Weak-coupling evaluation}
\label{sec:weak}
Suppose we regard $\lambda$ as a small parameter, so let us expand 
(\ref{energy}) in powers of $\lambda$.  The first term is
\be
\mathcal{E}^{(1)}=-\frac\lambda{8\pi a^2}\sum_{m=-\infty}^\infty \int_0^\infty
dx\,x^2\,\frac{d}{dx}K_m(x)I_m(x).\label{1storder}
\ee
The addition theorem for the modified Bessel functions is
\be
K_0(kP)=\sum_{m=-\infty}^\infty e^{im(\phi-\phi')}K_m(k\rho)I_m(k\rho'),\quad
\rho>\rho',\label{addthm}
\ee
where $P=\sqrt{\rho^2+\rho^{\prime2}-2\rho\rho'\cos(\phi-\phi')}$. If this
is extrapolated to the limit $\rho'=\rho$ we conclude that the sum of the
Bessel functions appearing in (\ref{1storder}) is $K_0(0)$, that is, a 
constant, so there is no first-order contribution to the energy.
For a rigorous derivation of this result, see Sec.~\ref{sec:zeta1}
and also Appendix A.

\subsection{Analytic Regularization}
We can proceed the same way to evaluate the second-order contribution,
\be
\mathcal{E}^{(2)}=\frac{\lambda^2}{16\pi a^2}\int_0^\infty dx\,x^2\,
\frac{d}{dx}\sum_{m=-\infty}^\infty I_m^2(x)K_m^2(x).\label{2ndorden}
\ee
By squaring the sum rule (\ref{addthm}), and taking the limit
$\rho'\to\rho$, we evaluate the sum over Bessel functions appearing here
as
\be
\sum_{m=-\infty}^\infty I_m^2(x)K_m^2(x)=\int_0^{2\pi}\frac{d\varphi}{2\pi}
K_0^2(2x\sin\varphi/2).\label{squaresr}
\ee
Then changing the order of integration, the second-order energy can be
written as
\be
\mathcal{E}^{(2)}=-\frac{\lambda^2}{64\pi^2 a^2}\int_0^{2\pi}\frac{d\varphi}
{\sin^2\varphi/2}\int_0^\infty dz\,z\,K_0^2(z),\label{phive2}
\ee
where the Bessel-function integral has the value 1/2.  However, the integral
over $\varphi$ is divergent.  We interpret this integral by adopting
 an analytic regularization based on the integral \cite{Cavero-Pelaez:2004xp}
\be
\int_0^{2\pi}d\varphi \left(\sin\frac\varphi2\right)^s=\frac{2\sqrt{\pi}\Gamma
\left(\frac{1+s}2\right)}{\Gamma\left(1+\frac{s}2\right)},\label{sinint}
\ee
which holds for $\mbox{Re}\,s>-1$.  Taking the right-side of this equation
to define the $\varphi$ integral for all $s$, we conclude that the 
$\varphi$ integral in (\ref{phive2}), 
and hence the second-order energy $\mathcal{E}^{(2)}$, is zero.

\subsection{Numerical Evaluation}
\label{sec:num}
Given that the above argument evidently formally omits divergent terms, it may
be more satisfactory, as in \cite{Cavero-Pelaez:2004xp}, to offer a 
numerical evaluation of $\mathcal{E}^{(2)}$. (The corresponding argument
for $\mathcal{E}^{(1)}$ is given in Appendix A.) We can very efficiently do so 
using the uniform asymptotic expansions ($m\to\infty$):
\begin{subequations}\label{uae}
\bea
I_m(x)&\sim&\sqrt{\frac{t}{2\pi m}}e^{m\eta}\left(1+\sum_k\frac{u_k(t)}{m^k}
\right),\label{uaei}\\
K_m(x)&\sim&\sqrt{\frac{\pi t}{2 m}}e^{-m\eta}\left(1+\sum_k(-1)^k
\frac{u_k(t)}{m^k}\right),\label{uaek}
\eea
\end{subequations}
where $x=mz$, $t=1/\sqrt{1+z^2}$, and the value of $\eta$ is irrelevant here.
The polynomials in $t$ appearing in (\ref{uae}) are generated by
\bea
u_0(t)&=&1,\quad
u_k(t)=\frac12t^2(1-t^2)u'_{k-1}(t)+\frac18\int_0^t ds(1-5s^2)u_{k-1}(s).
\eea
Thus the asymptotic behavior of the product of Bessel functions appearing
in (\ref{2ndorden}) is
\be
I_m^2(x)K_m^2(x)\sim\frac{t^2}{4m^2}\left(1+\sum_{k=1}^\infty \frac{r_k(t)}
{m^{2k}}\right).\label{i2k2}
\ee
The first three polynomials occurring here are
\begin{subequations}\label{r}
\bea
r_1(t)&=&\frac{t^2}4(1-6t^2+5t^4),\label{r1}\\
r_2(t)&=&\frac{t^4}{16}(7-148t^2+554 t^4-708 t^6+295 t^8),\label{r2}\\
r_3(t)&=&\frac{t^6}{16}(36-1666t^2+13775t^4-44272t^6\nonumber\\
&&\quad\mbox{}+67162t^8-48510t^{10}
+13475 t^{12}).\label{r3}
\eea
\end{subequations}

We now write the second-order energy (\ref{2ndorden}) as
\bea
\mathcal{E}^{(2)}&=&-\frac{\lambda^2}{8\pi a^2}\Bigg\{\int_0^\infty
dx\,x\left[I_0^2(x)K_0^2(x)-\frac1{4(1+x^2)}\right]\nonumber\\
&&\quad\mbox{}-\frac14\lim_{s\to 0} \left( \frac 1 2 +
\sum_{m=1}^\infty m^{-s}\right)\int_0^\infty dz
\,z^{2-s}\frac{d}{dz}\frac1{1+z^2}\nonumber\\
&&\quad\mbox{}+
2\int_0^\infty dz\,z\frac{t^2}4\sum_{m=1}^\infty \sum_{k=1}^3 \frac{r_k(t)}
{m^{2k}}\nonumber\\
&&\quad\mbox{}+2\sum_{m=1}^\infty
\int_0^\infty dx\,x\left[I_m^2(x)K_m^2(x)-\frac{t^2}{4m^2}
\left(1+\sum_{k=1}^3\frac{r_k(t)}{m^{2k}}\right)\right]\Bigg\}.\label{num}
\eea
In the final integral $z=x/m$. The successive terms are evaluated as
\bea
\mathcal{E}^{(2)}&\approx&-\frac{\lambda^2}{8\pi a^2}\Bigg[
\frac14(\gamma+\ln4)-\frac14\ln2\pi-\frac{\zeta(2)}{48}
+\frac{7\zeta(4)}{1920}-\frac{31\zeta(6)}{16128}\nonumber\\
&&\quad+0.000864+0.000006\Bigg]=-\frac{\lambda^2}{8\pi a^2}(0.000000),
\eea
where  in the last term in (\ref{num}) only the $m=1$ and 2 terms are 
significant. Therefore, we have demonstrated numerically
that the energy in order $\lambda^2$ is zero to an accuracy of better than 
$10^{-6}$.

The astute reader will note that we used a standard, but possibly
questionable, analytic regularization in defining the second term in
(\ref{num}), where the initial sum and integral are only
defined for $1<s<2$, and then the result is continued to $s=0$.  
Alternatively, we could follow Ref.~\cite{Cavero-Pelaez:2004xp} and insert
there an exponential regulator in each integral of $e^{-x\delta}$, with
$\delta$ to be taken to zero at the end of the calculation.  For $m\ne0$
$x$ becomes $mz$, and then the sum on $m$ becomes
\be
\sum_{m=1}^\infty e^{-mz\delta}=\frac1{e^{z\delta}-1}.\label{sumonm}
\ee
Then when we carry out the integral over $z$ we obtain for that term
\be
\frac\pi{8\delta}-\frac14\ln2\pi.\label{cutoff1}
\ee
Thus we obtain the same finite part as above, but in addition an explicitly
divergent term
\be
\mathcal{E}^{(2)}_{\rm div}=-\frac{\lambda^2}{64 a^2\delta}.
\ee
If we think of the cutoff in terms of a vanishing proper time $\tau$,
$\delta=\tau/a$, this divergent term is proportional to $1/a$, so the
divergence in the energy goes like $L/a$, if $L$ is the (very large) length
of the cylinder.  This is of the form of the shape divergence encountered
in Ref.~\cite{Cavero-Pelaez:2004xp}.

\section{Zeta-Function Approach}
\label{sec:zf}

For the massless case, in the zeta-function scheme the
regularized total energy reads 
\be {\cal E} = - \frac 1
{8\pi a^2} a^{2s} \left( 1+s \left[ -1+2 \ln (2\mu) \right]\right)
\sum_{m=-\infty} ^\infty \int_0^\infty dx \,\, x^{2-2s}
\frac{d}{d x} \ln \left( 1 + \lambda K_m (x) I_m (x)
\right) ,\ee where $\mu$ is an arbitrary mass scale.
 This result can, for example, be taken from Ref.~\cite{scandurra};
see also (8.3.16) of Ref.~\cite{kirsten}. In order
to find the total energy one needs to find the analytic
continuation of this expression to $s=0$. Formally, in the $s\to0$
limit, this is exactly (\ref{energy}), so this is the zeta-function 
regularization of that result.
 
\subsection{$\mathcal{E}^{(1)}$}
\label{sec:zeta1}
The first order of the energy, in the weak-coupling expansion,
reads 
\bea {\cal E}^{(1)} &=& - \frac \lambda {8\pi a^2} a^{2s}
\left( 1+s \left[ -1+2 \ln (2\mu) \right]\right) \sum_{m=-\infty}
^\infty \int_0^\infty dx \,\, x^{2-2s}
\frac{d}{d x}
 K_m (x) I_m (x) \nn\\
 &=&- \frac \lambda {8\pi a^2} a^{2s} \left(
1+s \left[ -1+2 \ln (2\mu) \right]\right) \bigg\{
\int_0^\infty dx \,x^{2-2s} \frac d {d x} K_0
(x) I_0 (x) \nn\\
& & \quad\mbox{}+ 2\sum_{m=1} ^{\infty} m^{2-2s}
\int_0^\infty dz \,\, z^{2-2s} \frac{d}{d z}
 K_m (m z) I_m (m z)\bigg\} ,\label{1}\eea
where, for the purpose of the following argument, the $m=0$
contribution has been separated off. 
To be precise, let us mention that the $m=0$ contribution should be 
thought of as coming from a limit $M \to 0$ of the integral 
$\int_M^\infty$. This is what one actually obtains in the contour 
integral formalism; for full details see Ref.~\cite{kirsten}. 
With this in mind, (6.2) is analytic in the strip $1<\mbox{Re}\, s<3/2$. The
substitution $x= m z$ has been done in the last integral to
make the convergence properties of the series and the integral
better apparent when using the uniform asymptotics of the Bessel
functions (as noted above; see below).

We want to find the analytic continuation of this expression to
$s=0$. As it stands, the expression is not valid about $s=0$ and
manipulations need to be done. In order to evaluate it further,
one would like to interchange summation and integration and use the
addition theorem for the Bessel functions. From the uniform asymptotic
expansion of the Bessel functions ($m> 0$), 
\be K_m (m z) I_m
(m z) \sim \frac t {2 m} + \frac{t^3} {16 m^3} (1-6 t^2 +5
t^4) + {\cal O} \left(\frac 1 { m^5}\right),\label{2}\ee 
as we have seen already in (\ref{i2k2}) and (\ref{r1}),
it is clear that the resulting summation is
divergent at $s=0$; one can see that as it stands the summation is
convergent only for $\mbox{Re}\, s >1$. One therefore needs to subtract
and add as many asymptotic terms as needed to make the resulting
summation convergent. In this procedure the hope is always that
the asymptotic terms which potentially could contain singular
contributions at $s=0$ can be handled analytically. Subtracting
all asymptotic terms given above, one sees the resulting summation
is well defined at $s=0$. 

Let us define $\tEo$ without the prefactor in (\ref{1}), 
\be \tEo=\sum_{m=-\infty}^\infty \int_0^\infty dx \,\, x^{2-2 s}
\frac{d}{d x} K_m (x) I_m (x), \ee 
which, given the above remarks, we rewrite as 
\bea \tEo &=&
\int_0^\infty dx \,\, x^{2-2 s} \frac d {d x}
K_0 (x) I_0 (x)
\nn\\
& & \quad\mbox{}+2\sum_{m=1}^{\infty}m^{2-2s}
\int_0^\infty dz \,\, z^{2-2s} \frac d {d z}
\left[ \frac t {2 m } + \frac{t^3 (1-6 t^2 + 5 t^4)}{16 m^3}
\right] \nn\\
& &\quad\mbox{}+2\sum_{m=1}^{\infty}m^{2-2s}
\int_0^\infty dz \,\, z^{2-2s} \frac
d {d z} \left[ K_m (m z) I_m (m z) - \frac t {2 m} -
\frac {t^3 (1-6 t^2 + 5 t^4)}{16 m^3}
\right] .\nn\\
\label{3}\eea

We need to find the analytical continuation of this expression to
$s=0$. For the first term we find for $\frac12<\mbox{Re}\,s<1$
\be \int_0^\infty dx\,
x^{2-2s} \frac d {d x} K_0 (x) I_0 (x) =
-\frac{\Gamma (2-s) \Gamma (1-s) \Gamma \left( s-\frac 1 2
\right)} {2 \sqrt \pi \Gamma (s)}, \label{6.6}\ee 
where the right side vanishes when analytically continued to $s=0$.

For the middle, asymptotic, terms in (\ref{3})  we use for $1-\frac{n}2<
\mbox{Re}\,s<2$ 
\be\int_0^\infty dz\,z^{2-2s}
\frac d {d z} \frac 1 {(1+z^2)^{n/2}} = - \frac{
\Gamma (2-s) \Gamma \left( \frac n 2 + s-1 \right)}{\Gamma \left(
\frac n 2 \right) }\label{genint},\ee
 for $n$ an integer. The $m$-series then leads to
the zeta function of Riemann, and putting these remarks together
we first obtain 
\bea & &2\sum_{m=1}^{\infty}m^{2-2s}
\int_0^\infty dz \,\, z^{2-2s} \frac d {d z}
\left[ \frac t {2 m } + \frac{t^3 (1-6 t^2 + 5 t^4)}{16 m^3}
\right]\nn\\
&=&- \zeta _R (2s-1) \frac{\Gamma (2-s) \Gamma \left( s - \frac 1
2\right)} {\sqrt \pi} \nn\\
& &\quad\mbox{}+ \frac 1 8 \zeta _R (1+2s) \Gamma (2-s) \left\{ -\frac {\Gamma
\left( s + \frac 1 2 \right) } {\Gamma \left( \frac 3 2 \right) }
+ 6 \frac{ \Gamma \left( s + \frac 3 2 \right) } {\Gamma \left(
\frac 5 2 \right) } -5 \frac{ \Gamma \left( s + \frac 5 2 \right)
} { \Gamma \left( \frac 7 2 \right) } \right\} \nn\\
&=& - \zeta _ R (2s-1) \frac {\Gamma (2-s) \Gamma \left( s -
\frac 1 2 \right) } {\sqrt \pi} - \zeta _R (1+2s) \frac{ s (s-1)
\Gamma (2-s) \Gamma \left( s + \frac 1 2 \right) } {3 \sqrt \pi}.\label{6.8}
\eea
This derivation, which assumes $1<\mbox{Re}\,s<2$,
provides the analytic continuation to $s=0$. In
the second term one needs to be a little careful as the pole at
$s=0$ coming from the $\zeta_R (1+2s)$ multiplies $s$,
contributing altogether something finite. Adding up the two
contributions, namely $-1/6$ and $+1/6$, the total contribution is
$0$.

This shows that $m=0$ and the asymptotic terms do not contribute
to order $\lambda$.  The fact that the formulas (\ref{6.6}) and (\ref{6.8})
have non-overlapping domains of validity is irrelevant,
taking into account the remark made below (\ref{1}).

Let us now consider the most difficult term (which usually one is
not able to handle analytically, but here it works out fine). We
want to find 
\be Z ^{(1)} = \int_0 ^\infty dx \,\, x^2
\frac d {d x} 2\sum_{m=1} ^ {\infty}
\left\{ K_m (x) I_m (x) - \frac t {2 m } - \frac{
t^3 (1-6 t^2 + 5t^4) }{16 m^3} \right\},\ee 
which is the
last term in (\ref{3}) at $s=0$ with the substitution $x=z m$
and therefore here $t=1/\sqrt{1+(x/m)^2}$; this representation, at
least for the first term, is better for the following application
of the addition theorem of the Bessel functions. Remember, we
are able to put $s=0$ and interchange summation and integration because
$Z^{(1)}$ is well defined by construction.

We will now show that this finite quantity actually vanishes.

As it stands, the summation cannot be performed. In order to be
able to deal with the individual terms separately, we introduce an
oscillatory factor. For $Z^{(1)}$ we therefore write 
\be Z ^{(1)}
=\lim_{\varphi \to 0}  \int_0 ^\infty dx \,\, x^2 \frac
d {d x} 2\mbox{Re}\,\sum_{m=1} ^ {\infty} e^{i
m\varphi} \left\{ K_m (x) I_m (x) - \frac t {2 m } - \frac{ t^3
(1-6 t^2 + 5t^4) }{16 m^3} \right\}. \ee
The advantage is
that we can now use the addition theorem for Bessel functions
(\ref{addthm}) to see that
\bea 2\mbox{Re}\,\sum_{m=1} ^{\infty} e^{im\varphi} K_m
(x) I_m (x) = K_0 \left( 2 x \sin \frac \varphi 2 \right) - K_0
(z) I_0 (z) .\eea 
This shows that 
\bea Z^{(1)} &=&
\lim_{\varphi \to 0} \int_0^\infty dx \,\, x^2 \frac
d {d x} \bigg\{ K_0\left( 2 x \sin \frac \varphi 2
\right) - K_0 (x) I_0 (x) \nn\\
& & - \sum_{m=1}^\infty \frac{\cos (m\varphi)} m t - \frac 1
8 \sum_{m=1} ^\infty \frac{\cos (m\varphi )} {m^3} t^3 (1-6 t^2 +
5 t^4 ) \bigg\}.\eea 
As they stand, the two series cannot be
performed further, because $t$ contains a nontrivial $m$-dependence. Of
course one would like to substitute back $x=z m$, but in order to
do so one needs to separate all terms and perform the
$x$-integration for each term individually. This gives divergent
results. So in order to be able to consider each term
individually, we have additionally to regularize the $x$-integration and will
therefore consider (effectively reinserting the zeta-function regularization)
\be Z^{(1)}_{\rm reg} (s) = \lim _{\varphi \to
0} \lim_{\alpha \to 0} \left[ Z_{11} (\alpha, \varphi ) + Z_{12}
(\alpha , \varphi ) + Z_{13} (\alpha , \varphi ) + Z_{14} (\alpha
, \varphi ) \right], \ee
 where 
 \begin{subequations}
 \bea Z_{11} (\alpha , \varphi )
&=& \int_0^\infty dx \,\, x^{2-\alpha} \frac d
{d x} K_0
\left( 2 x \sin \frac \varphi 2 \right), \\
Z_{12} (\alpha , \varphi ) &=& -\int_0^\infty dx \,\,
x^{2-\alpha }
\frac d {d x} K_0 (x) I_0 (x), \\
Z_{13} (\alpha , \varphi ) &=& - \int_0^\infty dx \,\,
x^{2-\alpha } \frac d {d x} \sum_{m=1}^\infty \frac{
\cos(m\varphi ) } m t, \\
Z_{14} (\alpha , \varphi ) &=& - \frac 1 8\int_0^\infty dx
\,\, x^{2-\alpha} \frac d {d x} \sum_{m=1}^\infty
\frac{ \cos (m\varphi )}{m^3} t^3 (1-6t^2 + 5 t^4 ). \eea
\end{subequations}
 The integral for $Z_{11}$ can be performed for $\mbox{Re}\,\alpha<2$ and one obtains 
 \be Z_{11}
(\alpha , \varphi) = - \frac 1 2 \Gamma \left( 1- \frac \alpha 2
\right) \Gamma \left( 2 - \frac \alpha 2 \right) \sin ^{ \alpha
-2} \left( \frac \varphi 2 \right).\ee 
At $\alpha =0$ this gives 
\be Z_{11} (0, \varphi ) = - \frac 1 2 \sin ^{-2} \left(
\frac \varphi 2 \right). \label{6.12}\ee 
Ultimately we want to send
$\varphi \to 0$; remembering that the oscillating factor was
introduced to make each separate sum convergent, it is expected that
in this limit divergences might occur. 
By construction, this divergent piece as $\varphi \to 0$ has to be
cancelled by one of the remaining contributions, since at the beginning
the combination of all terms was finite.

The second contribution is calculated as in (\ref{6.6}), namely for
$1<\mbox{Re}\,\alpha<2$,
\be
Z_{12} (\alpha , \varphi ) = (2-\alpha ) \frac {\Gamma \left( 1-
\frac \alpha 2 \right)^2 \Gamma \left( \frac{\alpha -1} 2 \right)}
{ 4 \sqrt \pi \Gamma \left( \frac \alpha 2 \right) },\ee 
which vanishes at $\alpha =0$. So this does not contribute to the
energy.

Next we find (we substitute back to the case where $t=1/\sqrt{1+z^2}$)
for $1<\mbox{Re}\,\alpha<4$ 
\be Z_{13} (\alpha , \varphi ) = - \left(
\int_0^\infty dz\, z^{2-\alpha} \frac d {d z} t
\right) \left( \sum_{m=1}^\infty m^{1-\alpha} \cos (m\varphi )
\right), \ee 
which at $\alpha =0$ gives 
\be Z_{13} (0, \varphi
) = \frac1{2\sin^2\varphi/2},\ee
which exactly cancels (\ref{6.12}).

Finally, we find for $-1<\mbox{Re}\,\alpha<4$ 
\be Z_{14} (\alpha , \varphi ) = - \frac
1 8 \left( \int_0^\infty dz \,\, z^{2-\alpha } \frac
d {d z} t^3 \left( 1- 6t^2 + 5t^4 \right) \right) \left(
\sum_{m=1} ^\infty \frac{ \cos (m\varphi ) } {m^{1+\alpha}}\right).
\ee 
Looking at the calculation of the asymptotic terms, it is
seen that the integral is the same as that encountered in the second
term in (\ref{6.8}) and hence at
$\alpha =0$ this vanishes. Note, that this time the $0$ is not
multiplied by any infinite quantity because $\varphi \neq 0$ at
this stage. So no contribution results from this term.

Adding up all the contributions, the answer is simply
$\tEo = 0$ and therefore 
\be{\cal E}^{(1)} = 0,\ee
in agreement with the argument given at the beginning of Sec.~\ref{sec:weak}.
(See also Appendix A.)

\subsection{$\mathcal{E}^{(2)}$}
The second order contribution, fortunately, is significantly
simpler. The zeta-regularized version of the second-order energy reads
\be \mathcal{E}^{(2)} = \lambda^2 \frac{a^{2s}}{16 \pi a^2} \left( 1 + s
\left[ -1 + 2 \ln (2\mu)\right] \right) \sum_{m=-\infty} ^\infty
\int_0^\infty dx \,\, x^{2-2s} \frac d {d x}
K_m ^2 (x) I_m^2 (x).\ee
As before, let us define the second
order term without the prefactor, 
\be \tEt =\sum_{m=-\infty} ^\infty
\int_0^\infty dx \,\, x^{2-2s} \frac d {d x}
K_m ^2 (x) I_m^2 (x).\label{et2}\ee 
As it stands, the representation is
well defined in the strip $1/2 < \mbox{Re}\, s <1 $. In that strip we
interchange integral and summation using (\ref{squaresr}).
Substituting $u=2x \sin \varphi /2$, we obtain for (\ref{et2})
\be \tEt
= \int_0 ^{2\pi } \frac {d \varphi } {2\pi} \left( 2 \sin
\frac \varphi 2 \right) ^{-2+2s} \int_0^\infty du \,\,
u^{2-2s } \frac d {d u} K_0^2 (u). \ee 
In the
relevant range, namely $1/2 < \mbox{Re}\, s <1$, we have the integral 
[(\ref{sinint})]
\be
\int_0^{2\pi} d\varphi \left( \sin \frac \varphi 2 \right)
^{-2+2s} = \frac {2 \sqrt \pi \Gamma \left( s- \frac 1 2 \right) }
{\Gamma (s)} .\ee 
So the analytic continuation of the
integral to $s=0$ vanishes. Given that the $u$-integral is well behaved
about $s=0$ (equaling $-1$ there), 
we have already found the final answer to be $\tEt = 0 $ and thus 
\be{\cal E}^{(2)} =0.\ee The first two
orders of the weak-coupling expansion indeed do identically vanish.

\section{Divergences in the total energy}
\label{hk}
In the following we are going to use heat-kernel knowledge to
determine the divergence structure in the total energy. We
consider a general cylinder of the type ${\cal C} = \mathbb{R} \times
Y$, where $Y$ is an arbitrary smooth two dimensional region rather
than merely being the disc. As a metric we have $ds^2 = dz^2 +
dY^2$ from which we obtain that the zeta function (density)
associated with the Laplacian on ${\cal C}$ is ($\mbox{Re}\,s>3/2$) 
\bea
 \zeta (s) &=& \frac 1 {2\pi} \int_{-\infty}^\infty dk
 \sum_{\lambda_Y} (k^2 + \lambda_Y)^{-s} = \frac 1 {2\pi}
 \frac{\sqrt \pi \Gamma \left( s-\frac 1 2 \right)}{\Gamma (s)}
 \sum _{\lambda_Y} \lambda_Y ^{1/2 -s} \nn\\
&=& \frac 1 {2\pi}
 \frac{\sqrt \pi \Gamma \left( s-\frac 1 2 \right)}{\Gamma (s)} \zeta_Y 
\left(s-\frac 1 2 \right) .\eea
Here $\lambda_Y$ are the eigenvalues of the Laplacian on $Y$, and
$\zeta_Y (s)$ is the zeta function associated with these
eigenvalues. In the zeta-function scheme the Casimir energy is
defined as 
\be \left. E_{\rm Cas} = \frac 1 2 \mu^{2s} \,\,\zeta
\left( s-\frac 1 2 \right) \right|_{s=0}, \ee
which, in the
present setting, turns into 
\bea \left. E_{\rm Cas} = \frac 1 2
\mu^{2s} \frac {\Gamma (s-1)} {2\sqrt \pi \Gamma \left( s- \frac 1
2 \right)} \zeta_Y (s-1) \right|_{s=0} \eea
Expanding this
expression about $s=0$, one obtains 
\be E_{\rm Cas} = \frac 1 {8\pi
s} \zeta_Y (-1) + \frac 1 {8\pi} \left( \zeta _Y (-1) \left[ 2 \ln
(2\mu ) -1\right] + \zeta_Y ' (-1) \right) + {\cal O } (s).\ee
The contribution associated with $\zeta _Y (-1)$ can be
determined solely from the heat-kernel coefficient knowledge,
namely 
\be \zeta _Y (-1) = - a_4, \ee
in terms of the standard 4th heat-kernel coefficient. 
The contribution coming
from $\zeta _Y ' (-1)$ can in general not be determined.
But as we see, at least the divergent term can be determined entirely by
the heat-kernel coefficient. 

The situation considered
in the Casimir energy calculation is a $\delta$-function shell
along some smooth line $\Sigma$ in the plane (here, a
circle of radius $a$). The manifolds considered
are the cylinder created by the region inside of the line, and
the region outside of the line; from the results the
contribution from free Minkowski space has to be subtracted to avoid
trivial volume divergences (the representation in terms of the
Bessel functions already has Minkowski space contributions
subtracted). The $\delta$-function shell generates a jump in the
normal derivative of the eigenfunctions; 
call the jump $U$ (here, $U=\lambda/a$). 
The leading heat-kernel coefficients for
this situation, namely for functions which are continuous 
across the boundary but which have
a jump of the first normal derivative at the boundary, have been
determined in Ref.~\cite{gilkey}; the
relevant $a_4$ coefficient is given in Theorem 7.1, p.~139 of that
reference. The
results there are very general; for our purpose there is exactly one
term that survives, namely 
\be a_4 = - \frac 1 {24 \pi}
\int_\Sigma dl\, U^3 \ee
 which shows that
\be E_{\rm Cas}^{\rm div} =
\frac 1 {192 \pi^2 s}\int_\Sigma dl\, U^3 \ee
So no matter
along which line the $\delta$-function shell is concentrated, the
first two orders in a weak-coupling expansion do not contribute
any divergences in the total energy. 
But the third order does, and the divergence is given above.

For the example considered, as mentioned, $U=\lambda / a$ is
constant, and the integration leads to the length of the line
which is $2\pi a$. Thus we get for this particular
example 
\be \mathcal{E}_{\rm Cas}^{\rm div} = \frac 1 {96 \pi s}
\frac{\lambda^3}{a^2} \label{lambda3div}
\ee 
This can be easily checked from the
explicit representation we have for the energy. We have already
seen that the first two orders in $\lambda$ identically vanish, while
the part of the third order that potentially contributes a
divergent piece is 
\be - \frac 1 {8 \pi a^2} \sum_{m=-\infty}
^\infty \int_0^\infty dx\, x^{2-2s} \frac d {d
x} \frac 1 3\lambda^3 K_m ^3 (x) I_m ^3 (x).\label{pdp} \ee 
The $m=0$ contribution is well behaved about $s=0$; while for the remaining sum
using 
\be K_m^3 (mz) I_m ^3 (mz) \sim \frac 1 {8m^3} \frac 1
{(1+z^2)^{3/2} },\ee
 we see that the leading contribution is from (\ref{genint}) 
\bea
&&-\frac{\lambda^3}{12 \pi a^2} \sum_{m=1}^\infty m^{2-2s}
\int_0^\infty dz \,z^{2-2s} \frac d {d z} \frac
1 {8m^3} \frac 1 {(1+z^2)^{3/2}} \nn\\
&=& - \frac {\lambda^3}{96 \pi
a^2} \zeta_R (1+2s) \int_0^\infty dz z^{2-2s} \frac
d {d z} \frac 1 {(1+z^2)^{3/2}} \nn\\
& =&\frac {\lambda^3}{96 \pi a^2} \zeta_R
(1+2s)\frac{\Gamma (2-s) \Gamma \left( s+\frac 1 2 \right)}
{\Gamma (3/2)} = \frac {\lambda^3}{96 \pi a^2 s} + {\cal O} (s^0),
\eea in perfect agreement with the heat-kernel prediction (\ref{lambda3div}).

\section{Strong Coupling}
\label{sec:sc}
The strong-coupling limit of the energy (\ref{energy}), that is, the
Casimir energy of a Dirichlet cylinder, 
\be
\mathcal{E}^D=-\frac1{8\pi a^2}\sum_{m=-\infty}^\infty \int_0^\infty dx\,x^2
\frac{d}{dx}\ln I_m(x)K_m(x),
\ee
was worked out to high accuracy by Gosdzinsky and Romeo \cite{gosrom},
\be
\mathcal{E}^D=\frac{0.000614794033}{a^2}.\label{grresult}
\ee
It was later redone with less accuracy by Nesterenko and Pirozhenko
\cite{nest}.  (Note that the preprint version of the latter contains
several internal sign errors.)

For completeness, let us sketch the evaluation here.  
We carry out a numerical calculation (very similar to that of \cite{nest})
in the spirit of Sec.~\ref{sec:num}.  We add and subtract the leading
uniform asymptotic expansion (for $m=0$ the asymptotic behavior) as follows:
\bea
\mathcal{E}^D&=&-\frac1{8\pi a^2}\Bigg\{-2\int_0^\infty dx\,x\left[
\ln\left(2xI_0(x)K_0(x)\right)-\frac18\frac1{1+x^2}\right]\nonumber\\
&&\quad\mbox{}+2\sum_{m=1}^\infty \int_0^\infty dx\,x^2\frac{d}{dx}
\left[\ln \left(2x I_m(x)K_m(x)
\right)-\ln\left(\frac{xt}{m}\right)-\frac12\frac{r_1(t)}{m^2}\right]
\nonumber\\
&&\quad\mbox{}-2\left(\frac12+\sum_{m=1}^\infty\right)
\int_0^\infty dx\,x^2\frac{d}{dx}\ln 2x
+2\sum_{m=1}^\infty \int_0^\infty dx\,x^2 \frac{d}{dx}\ln xt
\nonumber\\
&&\quad\mbox{}+\sum_{m=1}^\infty \int_0^\infty dx\,x^2 \frac{d}{dx}\left[
\frac{r_1(t)}{m^2}-\frac14\frac1{1+x^2}\right]\nonumber\\
&&\quad\mbox{}+\frac14\left(\frac12+\sum_{m=1}^\infty\right)
\int_0^\infty dx\,x^2\frac{d}{dx}\frac{1}{1+x^2}\Bigg\}.
\label{scints}\eea
In the first two terms we have subtracted the leading asymptotic behavior so
the resulting integrals are convergent.  Those terms are restored in the 
fourth, fifth, and sixth terms.  The most divergent part of the Bessel 
functions
are removed by the insertion of $2x$ in the corresponding integral, and
its removal in the third term.  (Elsewhere, such terms have been referred to
as ``contact terms.'')  The terms involving Bessel functions are evaluated
numerically, where it is observed that the asymptotic value of the
summand (for large $m$)
in the second term is $1/32m^2$.  The fourth term is evaluated by writing it
as
\be
2\lim_{s\to0}\sum_{m=1}^\infty m^{2-s}\int_0^\infty dz
\frac{z^{1-s}}{1+z^2}=2\zeta'(-2)=-\frac{\zeta(3)}{2\pi^2},\label{4th}
\ee
while the same argument, as anticipated, shows that the third ``contact'' term
is zero,\footnote{This argument is a bit suspect, since the
analytic continuation that defines the integrals has no common region
of existence.  Thus the argument in the following subsection may be
preferable.} while the sixth term is
\be
-\frac12\lim_{s\to 0}\left[\zeta(s)+\frac12\right]\frac1s=\frac14\ln 2\pi.
\ee
The fifth term is elementary.  The result then is
\bea
\mathcal{E}^D&=&\frac1{4\pi a^2}\left(0.010963-0.0227032+0+0.0304485+0.21875
-0.229735\right)\nonumber\\
&=&\frac1{4\pi a^2}(0.007724)=\frac{0.0006146}{a^2},\label{myresult}
\eea
which agrees with (\ref{grresult}) to the fourth significant figure.

\subsection{Exponential Regulator}
As in Sec.~\ref{sec:weak}, it may seem more satisfactory to insert an
exponential regulator rather than use analytic regularization.  Now it is
the third, fourth, and sixth terms in (\ref{scints}) that must be treated.
The latter is just the negative of (\ref{cutoff1}).  We can combine the
third and fourth terms to give using (\ref{sumonm})
\be
-\frac1{\delta^2}-\frac2{\delta^2}\int_0^\infty \frac{dz\,z^3}{z^2+\delta^2}
\frac{d^2}{dz^2}\frac{1}{e^z-1}.
\ee
The latter integral may be evaluated by writing it as an integral along
the entire $z$ axis, and closing the contour in the upper half plane,
thereby encircling the poles at $i\delta$ and at $2in\pi$, where $n$ is a
positive integer.  The residue theorem then gives for that integral
\be
-\frac{2\pi}{\delta^3}-\frac{\zeta(3)}{2\pi^2},
\ee 
so once again we obtain the same finite part as in (\ref{4th}).
In this way of proceeding, then, in addition to the finite part in 
(\ref{myresult}), we obtain divergent terms
\be
\mathcal{E}^D_{\rm div}=\frac1{64a^2\delta}+\frac1{8\pi a^2\delta^2}
+\frac1{4a^2\delta^3},
\ee
which, with the previous interpretation for $\delta$, implies divergent
terms in the energy proportional to $L/a$ (shape), $L$ (length), and
$aL$ (area), respectively.  Such terms presumably are to be subsumed
in a renormalization of parameters in the model.  Had a logarithmic
divergence occurred [as does occur in weak coupling in 
$\mathcal{O}(\lambda^3)$] such a
renormalization would apparently be impossible---however, see 
Sec.~\ref{sec:thick}.

\section{Local Energy Density}
\label{sec:local}
We compute the energy density from the stress tensor (\ref{st}), or
\bea
\langle T^{00}\rangle&=&\frac1{2i}\left(\partial^0\partial^{0\prime}
+\bm{\nabla}\cdot\bm{\nabla}'\right)G(x,x')\bigg|_{x'=x}-\frac\xi{i}\nabla^2G(x,x)
\nonumber\\
&=&\frac1{16\pi^3i}\int_{-\infty}^\infty dk\int_{-\infty}^\infty d\omega
\sum_{m=-\infty}^\infty \Bigg[\left(\omega^2+k^2+\frac{m^2}{r^2}+\partial_r
\partial_{r'}\right)g(r,r')\nonumber\\
&&\quad\mbox{}-2\xi\frac1r\partial_r r\partial_r g(r,r)\Bigg].
\eea
We omit the free part of the Green's function, since that corresponds to
the energy that would be present in the vacuum in the absence of
the cylinder. When we
insert the remainder of the Green's function (\ref{gout}), we obtain
the following expression for the energy density outside the cylindrical
shell:
\bea
u(r)&=&\langle T^{00}-T_{(0)}^{00}\rangle=-\frac\lambda{16\pi^3}
\int_{-\infty}^\infty
d\zeta\int_{-\infty}^\infty dk\sum_{m=-\infty}^\infty \frac{
I_m^2(\kappa a)}{1+\lambda I_m(\kappa a)K_m(\kappa a)}
\nonumber\\
&&\times\left[\left(2\omega^2+\kappa^2+\frac{m^2}{r^2}\right)K_m^2(\kappa
r)+\kappa^2K_m^{\prime 2}(\kappa r)-2\xi\frac1r\frac\partial{\partial r}
r\frac\partial{\partial r} K_m^2(\kappa r)\right],\nonumber\\
&&\qquad\qquad r>a.\label{uofr}
\eea
The last factor in square brackets can be easily seen to be, from the
modified Bessel equation, 
\be
2\omega^2 K_m^2(\kappa r)+\frac{1-4\xi}2\frac1r\frac\partial{\partial r}
r\frac\partial{\partial r} K_m^2(\kappa r).\label{9.3}
\ee
For the interior region, $r<a$, we have the corresponding expression for
the energy density with $I_m\leftrightarrow K_m$.

\subsection{Total and Surface Energy}
We first need to verify that we recover the expression for the energy
found in Sec.~\ref{sec:energy}.  So let us integrate expression (\ref{uofr})
over the region exterior of the cylinder, and the corresponding
interior expression over the inside region. The second term in (\ref{9.3})
is a total derivative, while the first may be integrated
according to (\ref{int2}).  In fact that term is exactly the one evaluated in 
Sec.~\ref{sec:energy}.  The result is
\bea
2\pi\int_0^\infty dr\,r\,u(r)&=&-\frac1{8\pi a^2}\sum_{m=-\infty}^\infty
\int_0^\infty dx\,x^2\frac{d}{dx}
\ln\left[1+\lambda I_m(x)K_m(x)\right]\nonumber\\
&&\mbox{}-(1-4\xi)\frac\lambda{4\pi a^2}
\int_0^\infty dx\,x
\sum_{m=-\infty}^\infty \frac{I_m(x)K_m(x)}{1+\lambda I_m(x)K_m(x)}.\nonumber\\
\label{inten}
\eea
The first term is the total energy (\ref{energy}), but what do we make of
the second term?  In strong coupling, it would represent a constant that
should have no physical significance (a contact term---it is independent
of $a$ if we revert to the physical variable $\kappa$ as the integration
variable).  In general, however,
there is another contribution to the total energy, residing precisely
on the singular surface.  This surface energy is given in general
by \cite{Dowker:1978md,Kennedy:1979ar,saharian,fulling,milton04a}
\be
\mathfrak{E}=-\frac{1-4\xi}{2i}\oint_S d\mathbf{S}\cdot\bm{\nabla} G(x,x')
\bigg|_{x'=x},\label{surfen}
\ee
where the normal to the surface is out of the region in question.
In this case it is easy to see that $\mathfrak{E}$ exactly equals the
negative of the second term in (\ref{inten}).  This is an example of the
general theorem
\be
\int(d\mathbf{r}) u(\mathbf{r})+\mathfrak{E}=E,
\ee
that is, the total energy $E$
is the sum of the integrated local energy density
and the surface energy.  A consequence of this theorem is that the
total energy, unlike the local energy density, is independent of
the conformal parameter $\xi$.  For more on the surface term, see
Sec.~\ref{sec:surfenergy} and Appendix B.

\subsection{Surface Divergences}
We now turn to an examination of the behavior of the local energy
density (\ref{uofr}) as $r$ approaches $a$ from outside the cylinder.
To do this we use the uniform asymptotic expansion (\ref{uaei}), (\ref{uaek}),
where now we need to know that $d\eta/d z=1/zt$.  Let us begin by
considering the strong-coupling limit, a Dirichlet cylinder.  If we stop with 
only the leading asymptotic behavior, we obtain the expression
\bea
u(r)&\sim&-\frac1{8\pi^3}\int_0^\infty d\kappa\,\kappa\,2\sum_{m=1}^\infty
e^{-m\chi}\Bigg\{\left[-\kappa^2+(1-4\xi)\left(\kappa^2+\frac{m^2}{r^2}\right)
\right]\frac{\pi t}{2m}\nonumber\\
&&\qquad\mbox{}+(1-4\xi)\kappa^2\frac{\pi}{2mt}\frac1{z^2}
\Bigg\},\qquad (\lambda\to\infty),
\eea
where
\be
\chi=-2\left[\eta(z)-\eta\left(z\frac{a}r\right)\right],
\ee
and we have carried out the angular integral as in (\ref{angint}).
Here we ignore the difference between $r$ and $a$ except in the exponent, and
we now replace $\kappa$ by $m z/a$.  Close to the surface,
\be
\chi\sim \frac2t\frac{r-a}r,\quad r-a\ll r,
\ee
and we carry out the sum over $m$ according to
\be
2\sum_{m=1}^\infty m^3 e^{-m\chi}\sim-2\frac{d^3}{d\chi^3}\frac1\chi
=\frac{12}{\chi^4}\sim\frac34\frac{t^4r^4}{(r-a)^4}.
\ee
Then the energy density behaves, as $r\to a+$, 
\bea
u(r)&\sim&-\frac3{64\pi^2}\frac1{(r-a)^4}\int_0^\infty dz\,z[t^5+t^3(1-8\xi)]
\nonumber\\
&=&-\frac1{16\pi^2}\frac1{(r-a)^4}(1-6\xi).
\eea
This is the universal surface divergence first discovered by Deutsch
and Candelas \cite{Deutsch:1978sc}.  It therefore occurs, with precisely the
same numerical coefficient, near a Dirichlet plate \cite{milton03}
or a Dirichlet sphere \cite{Cavero-Pelaez:15kq}.  It is utterly without physical
significance, and may be eliminated with the conformal choice for the
parameter $\xi$, $\xi=1/6$.

We will henceforth make this conformal choice.  Then the leading divergence
depends upon the curvature.  This was also worked out by Deutsch and
Candelas \cite{Deutsch:1978sc}; for the case of a cylinder, that result is
\be
u(r)\sim \frac1{720\pi^2}\frac1{r(r-a)^3},\quad r\to a+,\label{dccyl}
\ee
exactly 1/2 that for a Dirichlet sphere of radius $a$ 
\cite{Cavero-Pelaez:15kq}.
Here, this result may be straightforwardly derived by keeping the
$1/m$ corrections in the uniform asymptotic expansion (\ref{uaei}), 
(\ref{uaek}), as well as the next term in the expansion of $\chi$,
\be
\chi\sim\frac2t\frac{r-a}r+t\left(\frac{r-a}r\right)^2.\label{chiexp}
\ee
(Note that there is a sign error in (4.8) in Ref.~\cite{Cavero-Pelaez:15kq}.)

\subsection{Weak Coupling}

Let us now expand the energy density (\ref{uofr}) for small coupling,
\bea
u(r)&=&-\frac{\lambda}{16\pi^3}\int_{-\infty}^\infty d\zeta
\int_{-\infty}^\infty dk\sum_{m=-\infty}^\infty I_m^2(\kappa a)
\sum_{n=0}^\infty(-\lambda)^n I_m^n(\kappa a)K_m^n(\kappa a)\nonumber\\
&&\quad\times\left\{\left[-\kappa^2+(1-4\xi)\left(\kappa^2+\frac{m^2}{r^2}
\right)
\right]K_m^2(\kappa r)+(1-4\xi)\kappa^2 K_m^{\prime2}(\kappa r)\right\}.
\eea
If we again use the leading uniform asymptotic expansions for the Bessel
functions, we obtain the expression for the leading behavior of the term 
of order $\lambda^{n}$,
\be
u^{(n)}(r)\sim \frac1{8\pi^2r^4}\left(-\frac\lambda2\right)^{n}
\int_0^\infty dz\,z\sum_{m=1}^\infty m^{3-n}e^{-m\chi}t^{n-1}(t^2+1-8\xi).
\ee
The sum on $m$ is asymptotic to
\be
\sum_{m=1}^\infty m^{3-n}e^{-m\chi}\sim (3-n)!\left(\frac{t r}{2(r-a)}
\right)^{4-n},\quad r\to a+,\label{summ}
\ee
so the most singular behavior of the order $\lambda^n$ term is, as $r\to a+$,
\be
u^{(n)}(r)\sim (-\lambda)^n\frac{(3-n)!\,(1-6\xi)}{96\pi^2 r^n(r-a)^{4-n}}.
\ee
This is exactly the result found for the weak-coupling limit for a 
$\delta$-sphere \cite{Cavero-Pelaez:15kq} and for a $\delta$-plane \cite{milton04a},
so this is also a universal result, without physical significance.  It may be
made to vanish by choosing the conformal value $\xi=1/6$.

With this conformal choice, once again we must expand to higher order.
Besides the corrections noted above, in (\ref{uaei}), (\ref{uaek}), and
(\ref{chiexp}), we also need
\be
\tilde t\equiv t(z a/r)\sim t+(t-t^3)\frac{r-a}r,\qquad r\to a,
\ee
Then a quite simple calculation gives
\be
u^{(n)}\sim(-\lambda)^n\frac{(n-1)(n+2)\Gamma(3-n)}{2880\pi^2 r^{n+1}
(r-a)^{3-n}},\quad r\to a_+,
\ee
which is analytically continued from the region $1\le \mbox{Re}\,
n<3$.  Remarkably, this
is exactly one-half the result found in the same weak-coupling expansion
for the leading conformal divergence outside a sphere \cite{Cavero-Pelaez:15kq}.
Therefore, like the strong-coupling result (\ref{dccyl}), 
this limit is universal, depending on the sum of the principal curvatures 
of the interface.

\section{Cylindrical Shell of Finite Thickness}
\label{sec:thick}
In this section we regard the shell (annulus) to have a finite
thickness $\delta$.  We consider the potential
\be
\mathcal{L}_{\rm int}=-\frac\lambda{2a}\phi^2\sigma(r),
\ee
where
\be
\sigma(r)=\left\{\begin{array}{cc}
0,&r<a_-,\\
h,&a_-<r<a_+,\\
0,&a_+<r.\end{array}\right.
\ee
Here $a_\pm=a\pm\delta/2$, and we set $h\delta=1$.  In the limit as
$\delta\to 0$ we recover the $\delta$-function potential considered in the
rest of this paper.

As in Ref.~\cite{Cavero-Pelaez:15kq} it is straightforward to find the
Green's function for this potential.  In fact, the result may be obtained
from the reduced Green's function given there by an evident substitution.
Here, we content ourselves by stating the result for the Green's function
in the region of the annulus, $a_-<r,r'<a_+$:
\bea
g_m(r,r')&=&I_m(\kappa'r_<)K_m(\kappa'r_>)+AI_m(\kappa'r)I_m(\kappa'r')\nn\\
&&\quad\mbox{}+B[I_m(\kappa'r)K_m(\kappa'r')+K_m(\kappa'r)I_m(\kappa'r')]
+C K_m(\kappa'r)K_m(\kappa'r'),
\eea
where $\kappa'=\sqrt{\kappa^2+\lambda h/a}$.  The coefficients appearing
here are
\begin{subequations}
\bea
A&=&-\frac1\Xi[\kappa I_m'(\kappa a_-)K_m(\kappa' a_-)-\kappa'
I_m(\kappa a_-)K'_m(\kappa'a_-)]\nn\\
&&\quad\times[\kappa K_m'(\kappa a_+)K_m(\kappa' a_+)-\kappa'
K_m(\kappa a_+)K'_m(\kappa'a_+)],\\
B&=&\frac1\Xi[\kappa I_m'(\kappa a_-)I_m(\kappa' a_-)-\kappa'
I_m(\kappa a_-)I'_m(\kappa'a_-)]\nn\\
&&\quad\times[\kappa K_m'(\kappa a_+)K_m(\kappa' a_+)-\kappa'
K_m(\kappa a_+)K'_m(\kappa'a_+)],\\
C&=&-\frac1\Xi[\kappa I_m'(\kappa a_-)I_m(\kappa' a_-)-\kappa'
I_m(\kappa a_-)I'_m(\kappa'a_-)]\nn\\
&&\quad\times[\kappa K_m'(\kappa a_+)I_m(\kappa' a_+)-\kappa'
K_m(\kappa a_+)I'_m(\kappa'a_+)],
\eea
\end{subequations}
where the denominator is
\bea\Xi&=&[\kappa I_m'(\kappa a_-)K_m(\kappa' a_-)-\kappa'
I_m(\kappa a_-)K'_m(\kappa'a_-)]\nn\\
&&\quad\times[\kappa K_m'(\kappa a_+)I_m(\kappa' a_+)-\kappa'
K_m(\kappa a_+)I'_m(\kappa'a_+)]\nn\\
&&\quad\mbox{}-[\kappa I_m'(\kappa a_-)I_m(\kappa' a_-)-\kappa'
I_m(\kappa a_-)I'_m(\kappa'a_-)]\nn\\
&&\quad\times[\kappa K_m'(\kappa a_+)K_m(\kappa' a_+)-\kappa'
K_m(\kappa a_+)K'_m(\kappa'a_+)].
\eea
The general expression for the energy
density within the shell is given in terms of these coefficients by
\bea
u&=&\frac1{8\pi^2}\int_0^\infty d\kappa \,\kappa\sum_{m=-\infty}^\infty
\bigg\{\left[-\frac{\kappa^2}2+\frac{m^2}{r^2}+k^2+\frac{\lambda h}a
-4\xi\kappa^{\prime 2}\left(1+\frac{m^2}{\kappa^{\prime 2}r^2}\right)
\right]\nn\\
&&\quad\times[A I_m^2(\kappa' r)+C K_m^2(\kappa'r)+2B K_m(\kappa'r)
I_m(\kappa'r)]\nn\\
&&\quad\mbox{}+\kappa^{\prime 2}(1-4\xi)
[A I_m^{\prime2}(\kappa' r)+C K_m^{\prime2}(\kappa'r)+2B K'_m(\kappa'r)
I'_m(\kappa'r)]\bigg\}\nn\\
&=&\frac1{8\pi^2}\int_0^\infty d\kappa\,\kappa\left[-\kappa^2+\frac{1-4\xi}2
\frac1r\frac\partial{\partial r}r\frac\partial{\partial r}\right]\nn\\
&&\quad\times\sum_{m=-\infty}^\infty
[A I_m^2(\kappa' r)+C K_m^2(\kappa'r)+2B K_m(\kappa'r)
I_m(\kappa'r)].\label{enshell}
\eea

\subsection{Leading Surface Divergence}
The above expressions are somewhat formidable.  Therefore, to isolate
the most divergent structure, we replace the Bessel functions by the leading
uniform asymptotic behavior (\ref{uae}).  A simple calculation implies
\begin{subequations}
\bea
A&\sim&\frac{t_+-t_+'}{t_++t_+'}e^{-2m\eta'_+},\\
B&\sim&\frac{t_+-t_+'}{t_++t_+'}\frac{t_--t_-'}{t_-+t_-'}e^{2m(\eta'_--
\eta'_+)},\\
C&\sim&\frac{t_--t_-'}{t_-+t_-'}e^{2m\eta'_-},
\eea
\end{subequations}
using the notation $t_+=t(z_+)$, $\eta'_-=\eta(z'_-)$, etc., where for example
$z_-'=\kappa'a_-/m$.  If we now insert this approximation into the
form (\ref{enshell}) for the energy density, we find
\bea
u&=&\langle T^{00}\rangle=\frac1{8\pi^2a_+^4}2\sum_{m=1}^\infty m\int_0^\infty
dz_+\,z_+ t_r'\nn\\
&&\quad\times\bigg\{\left[\frac{t_+-t_+'}{t_++t_+'}e^{2m(\eta_r'-\eta_+')}
+\frac{t_--t_-'}{t_-+t_-'}e^{2m(-\eta'_r+\eta'_-)}\right]\nn\\
&&\qquad\times\left[\frac{m^2z_+^2}2(1-8\xi)+\left(
\frac{\lambda ha_+^2}a+\frac{m^2a_+^2}{r^2}
\right)(1-4\xi)\right]\nn\\
&&\quad\mbox{}-m^2z_+^2\frac{t_+-t_+'}{t_++t_+'}\frac{t_--t_-'}{t_-+t_-'}
e^{2m(\eta'_--\eta'_+)}\bigg\}.
\eea

If we are interested in the surface divergence as $r$ approaches the outer
radius $a_+$, the dominant term comes from the first exponential factor
only.  Because we are considering the limit $\lambda h a\ll m^2$, we
have
\be
t_+'\approx t_+\left(1-\frac{\lambda h}{2m^2}\frac{a_+^2}a t_+^2\right),
\ee and we have
\be
u\sim-\frac{\lambda h /a}{32 \pi^2 a_+^2}\sum_{m=1}^\infty m\int_0^\infty
dz\,z t(1-8\xi +t^2)e^{2m(\eta_r'-\eta_+')}.
\ee
The sum over $m$ is carried out according to (\ref{summ}), or
\be
\sum_{m=1}^\infty m e^{2m(\eta'_r-\eta_+')}\sim\left(\frac{r t_r'}{2(r-a_+)}
\right)^2, 
\ee
and the remaining integrals over $z$ are elementary.  The result
is
\be
u\sim \frac{\lambda h}{96\pi^2a}\frac{1-6\xi}{(r-a_+)^2},
\quad r\to a_+.
\ee
This is the expected universal divergence of a scalar field near a
surface of discontinuity \cite{Bordag:1996zb}, which is without significance, 
and which may once again be eliminated by setting $\xi=1/6$.

\subsection{Surface Energy}
\label{sec:surfenergy}
In this subsection we want to establish that the surface energy $\mathfrak{E}$
(\ref{surfen}) is the same as the integrated local energy density in the
shell when the limit $\delta\to 0$ is taken.  To examine this limit,
we consider $\lambda h/a\gg \kappa^2$.  So we apply the uniform
asymptotic expansion for the Bessel functions of $\kappa'$ only. We must
keep the first two terms in powers of $\kappa\ll\kappa'$:
\bea
\Xi&\sim&-\kappa^{\prime2}\frac{I_m(\kappa a_-)K_m(\kappa a_+)}{mz_-'z_+'
\sqrt{t_-'t_+'}}\sinh m(\eta_-'-\eta_+')\nn\\
&&-\frac{\kappa'\kappa}m\left[\frac1{z_+'}\sqrt{\frac{t_-'}{t_+'}}I_m'(\kappa
a_-)K_m(\kappa a_+)-\frac1{z_-'}\sqrt{\frac{t_+'}{t_-'}}I_m(\kappa
a_-)K'_m(\kappa a_+)\right]\cosh m(\eta'_--\eta_+').\nn\\
\eea
Because we are now regarding the shell as very thin,
\be
\eta'_--\eta'_+\approx -\frac\delta a\frac1{t'},
\ee
where
\be
t'\sim \frac1{z'}\sim \frac{m}{\sqrt{\lambda h a}},
\ee using the Wronskian (\ref{wronskian}) we get
\be
\Xi\sim -\frac1{a^2}[1+\lambda I_m(\kappa a)K_m(\kappa a)].
\ee
Then we immediately find the interior coefficients:
\begin{subequations}
\bea
A&\sim&\frac\pi2\sqrt{\lambda h a}\frac{I_m(\kappa a)K_m(\kappa a)}{1+\lambda
I_m(\kappa a)K_m(\kappa a)}e^{-2m\eta'},\\
B&\sim&\frac12\sqrt{\lambda h a}\frac{I_m(\kappa a)K_m(\kappa a)}{1+\lambda
I_m(\kappa a)K_m(\kappa a)},\\
C&\sim&\frac1{2\pi}\sqrt{\lambda h a}\frac{I_m(\kappa a)K_m(\kappa a)}{1+\lambda
I_m(\kappa a)K_m(\kappa a)}e^{2m\eta'},
\eea
\end{subequations}

We now insert this in (\ref{enshell}) and
keep only the largest terms, thereby neglecting $\kappa^2$ relative to 
$\lambda h/a$.  This gives a leading term proportional to $h$, which
when multiplied by the area of the annulus $2\pi a\delta$ gives for
the energy in the shell
\be
\mathcal{E}_{\rm ann}=2\pi\delta a u\sim(1-4\xi)\frac{\lambda}{4\pi a^2}
\sum_{m=-\infty}^\infty \int_0^\infty d\kappa a\,\kappa a
\frac{I_m(\kappa a)K_m(\kappa a)}{1+\lambda I_m(\kappa a)K_m(\kappa a)}=
\mathfrak{E},\label{shellissurf}
\ee
which is exactly the form of the surface energy given by the negative
of the second term in (\ref{inten}).  

In particular, note that the term in $\mathfrak{E}$ of order $\lambda^3$
is, for the conformal value $\xi=1/6$, exactly equal to that term in the
total energy $\mathcal{E}$ (\ref{energy}): [see (\ref{pdp})]
\be
\mathfrak{E}^{(3)}=\mathcal{E}^{(3)}.
\ee
This means that the divergence encountered in the global 
energy (\ref{lambda3div})
is exactly accounted for by the divergence in the surface energy, which
would seem to provide strong evidence in favor of the renormalizablity
of that divergence.

\section{Conclusions}

In this paper we have extended the considerations applied 
to a spherical geometry in Ref.~\cite{Cavero-Pelaez:15kq}
to a cylindrical one.  Results for the general
structure of the local and global energies are rather as expected, with
the leading conformal divergences in the local energy density 
as the surface is approached being reduced
from their spherical values by a factor of 1/2.  The only possibly 
surprising result is that the weak-coupling limit of the total Casimir energy
for a $\delta$-function cylindrical shell vanishes through order $\lambda^2$,
in agreement with other perturbative results for a cylinder, as seen in
Table \ref{tab1}.  We do not yet have a complete explanation of this.
Schaden \cite{Schaden:2006qg} has presented a semiclassical calculation
in terms of
periodic rays which sheds light of this general phenomenon.  His technique
closely reproduces Boyer's result for the spherical shell, while giving
a vanishing result to all orders for the cylinder.  This cannot be the
whole story, however, since Casimir energies for cylinders only vanish
through second order, and do not vanish for a perfectly conducting
cylinder.  Here, the weak-coupling Casimir energy diverges in $\mathcal{O}
(\lambda^3)$, which has its origin in the singular nature of the surface
energy.

\appendix
\section{Perturbative Divergences}
A conventional field theorist might be surprised that we are able to
extract finite (actually zero) values for the global energy in order $\lambda$
and $\lambda^2$, when the latter contributions would seem to correspond to
divergent Feynman graphs.  Thus the first-order energy is given by ($T$ is the
very large time interval characterizing the space-time volume under 
consideration)
\be
E^{(1)}=\frac{i}{2T}\frac\lambda{a}D_+(0)\int(dx)\sigma(x),
\ee
which involves the massless scalar propagator $D_+(x)$ at zero coordinate,
corresponding to equal field points.  The latter is ill-defined.  For example,
for our cylindrical $\delta$-shell geometry, we would compute from this the
energy per unit length
\be
\mathcal{E}^{(1)}=\frac{i\lambda}{4\pi}\frac1\epsilon,\quad \epsilon\to0+,
\ee
which, although divergent, is imaginary and independent of $a$. Such a 
contribution should be irrelevant.

Less divergent, but still ambiguous, is the second-order expression
\be
E^{(2)}=\frac{i}{4T}\left(\frac\lambda{a}\right)^2\int (dx)(dy)\sigma(x)D_+(x-y)
\sigma(y)D_+(y-x).
\ee
This may be evaluated by a method very similar to that employed in 
Ref.~\cite{milton03}.  If we work in $D+1$ dimensions, we obtain
\be
\mathcal{E}^{(2)}=-\frac{\lambda^2}4\frac{\Gamma\left(\frac{3-D}2\right)}
{(2\pi)^{(D-3)/2}}\frac1{a^{(D+1)/2}}\int_0^\infty dx\,x^{D-2}J_0^2(x)
\int_0^\infty du\,u^{(D-3)/2}(1-u)^{(D-3)/2}.\label{e2exp}
\ee
For $D>1$ the second integral may be evaluated 
as a beta function.  The integral
over the squared Bessel function is, for $1<\mbox{Re}\,D<2$,
\be
\int_0^\infty dx\,x^{D-2} J_0^2(x)=\frac{\Gamma\left(1-\frac{D}2\right)
\Gamma\left(\frac{D-1}2\right)}{2\sqrt{\pi}\Gamma^2\left(\frac{3-D}2\right)}.
\label{intb}
\ee
When this is inserted into (\ref{e2exp}) a factor of $\Gamma
\left((3-D)/2\right)$ remains in the denominator, so
the second-order energy evidently vanishes in three
dimensions.

We can also proceed directly in three dimensions, if we regulate the integral
using a proper time method as described in Ref.~\cite{milton03}.  If, as there,
we exclude a coincident field term (contact term), we are led to
\be
\mathcal{E}^{(2)}=\frac{\lambda^2\pi^2}4\frac{d}{d\alpha}\int_0^\infty dq
\, q^{2\alpha+1}J_0^2(qa)\bigg|_{\alpha=0}.
\ee
We encounter the same Bessel-function integral as in (\ref{intb}), so 
continuing
the integral from $-1<\mbox{Re}\,
\alpha<-1/2$ we again see that $\mathcal{E}^{(2)}=0$.
(This same continuation process gives the correct second-order energy for
a sphere: $E^{(2)}=\lambda^2/32\pi a$ for a dimensionless coupling constant
$\lambda$.)

In the balance of this appendix we sketch an evaluation of $\mathcal{E}^{(1)}$
along the lines of that given for $\mathcal{E}^{(2)}$ in Sec.~\ref{sec:num}.
We regulate the sum and integral by inserting an exponential cutoff, $\delta
\to 0+$:
\be
\mathcal{E}^{(1)}=-\frac{\lambda}{4\pi a^2}\left(\frac12+\sum_{m=1}^\infty
\right)\int_0^\infty dx\,x^2\frac{d}{dx}I_m(x)K_m(x) e^{-x\delta},
\ee
which we break up into five parts,
\be
\mathcal{E}^{(1)}=-\frac{\lambda}{8\pi a^2}(\mbox{I}+\mbox{II}+\mbox{III}+
\mbox{IV}+\mbox{V}).
\ee
The first term is the $m=0$ contribution, suitably subtracted to make it
convergent (so the convergence factor may be omitted),
\be
\mbox{I}=\int_0^\infty dx\,x^2\frac{d}{dx}\left[I_0(x)K_0(x)
-\frac1{2\sqrt{1+x^2}}\right]=-1.\label{i}
\ee
The second term is the above subtraction,
\be
\mbox{II}=\frac12\int_0^\infty dx\,x^2\left(\frac{d}{dx}\frac1{\sqrt{1+x^2}}
\right)e^{-x\delta}\sim-\frac1{2\delta}+1,\label{ii}
\ee
as may be verified by breaking the integral in two parts at $\Lambda$,
$1\ll\Lambda\ll1/\delta$.
The third term is the sum over the $m$th Bessel function with the two
leading asymptotic approximants (\ref{2}) subtracted:
\be
\mbox{III}=
2\sum_{m=1}^\infty \int_0^\infty dx\,x^2\frac{d}{dx}\left[I_m(x)K_m(x)
-\frac{t}{2m}\left(1+\frac{t^2}{8m^2}(1-6t^2+5t^4)\right)\right]=0.\label{iii}
\ee
Numerically, each term in the sum seems to be zero to machine accuracy.
This is verified by computing the higher-order terms in that expansion, 
in terms of the polynomials in (\ref{r}),
\bea
&&I_m(x)K_m(x)
-\frac{t}{2m}\left(1+\frac{t^2}{8m^2}(1-6t^2+5t^4)\right)\nn\\
&\sim&
\frac{t}{4m^5}\left[r_2(t)-\frac14 r_1^2(t)\right]+\frac{t}{4m^7}\left[r_3(t)
-\frac12 r_1(t)r_2(t)+\frac18r_1^3(t)\right]+\dots,
\eea
which terms are easily seen to integrate to zero.
The fourth term is the leading subtraction which appeared in the third term:
\be
\mbox{IV}=\sum_{m=1}^\infty m\int_0^\infty dz\,z^2
\left(\frac{d}{dz}t\right) e^{-mz\delta}.
\ee
If we first carry out the sum on $m$ we obtain
\bea
\mbox{IV}&=&
-\frac14\int_0^\infty dz\,z^3\frac1{(1+z^2)^{3/2}}\frac1{\sinh^2z\delta/2}
\nn\\
&\sim&-\frac1{\delta^2}+\frac1{2\delta}-\frac16,\label{iv}
\eea
as again may be easily verified by breaking up the integral.
The final term, if unregulated, is the form of infinity times zero:
\be
\mbox{V}=
\frac18\sum_{m=1}^\infty\frac1m\int_0^\infty dz\,z^2\frac{d}{dz}(t^3-6t^5+
5t^7)e^{-m z\delta}.
\ee
Here the sum on $m$ gives
\be
\sum_{m=1}^\infty \frac1m e^{-mz\delta}=-\ln\left(1-e^{-z\delta}\right),
\ee
and so we can write
\be
\mbox{V}=
\frac1{16}\frac{d}{d\alpha}\int_0^1 du\,(1-u)^\alpha u^{-2-\alpha}(u^{3/2}
-6u^{5/2}+5u^{7/2})\bigg|_{\alpha=0}=\frac16.\label{v}
\ee
Adding together (\ref{i}), (\ref{ii}), (\ref{iii}), (\ref{iv}), and (\ref{v}), 
we obtain 
\be
\mathcal{E}^{(1)}=\frac{\lambda}{8\pi a^2\delta^2}+0,
\ee
that is, the $1/\delta$ and constant terms cancel.  The remaining divergence
may be interpreted as an irrelevant constant, since $\delta=\tau/a$, $\tau$
being regarded as a point-splitting parameter.  The correspondence of the
terms in this evaluation with that Sec.~\ref{sec:zf} is rather immediate.

\section{Surface Energy}
Here, we suggest an alternative derivation of the surface energy term for
$\xi=0$.  For the scalar Lagrangian
\be
\mathcal{L}=-\frac12\partial_\mu\phi\partial^\mu\phi-\frac\lambda{2a}
\delta(r-a)\phi^2,
\ee
the response to a general coordinate transformation
yields the expected stress tensor, including the interaction term with
the cylindrical surface:
\be
T^{\mu\nu}=\partial^\mu\phi\partial^\nu\phi+g^{\mu\nu}\mathcal{L}.
\ee
The integral of the $\delta$-function term in the energy density should
be the surface energy
\be
\mathfrak{E}=2\pi\int_0^\infty dr\,r\,\frac\lambda{2a}\delta(r-a)
\langle\phi^2(r)\rangle=\pi\lambda\langle\phi(a)^2\rangle,
\ee
or, in terms of the Green's function
\bea
\mathfrak{E}&=&\frac{\pi\lambda}i\int_{-\infty}^\infty \frac{d\omega}{2\pi}
\int_{-\infty}^\infty \frac{dk}{2\pi}\frac1{2\pi}\sum_{m=-\infty}^\infty 
g_m(a,a)\nonumber\\
&=&-\frac\lambda{4\pi a^2}\int_0^\infty dx\,x\sum_{m=-\infty}^\infty
\frac{\lambda K_m^2(x)I_m^2(x)}
{1+\lambda I_m(x)K_m(x)}.
\eea
This is similar to the surface energy given by (\ref{shellissurf})
for $\xi=0$; the difference between the two expressions is
\be
-\frac\lambda{4\pi a^2}\sum_{m=-\infty}^\infty \int_0^\infty dx\,x\, I_m(x)K_m(x)
=-\frac\lambda{4\pi}\int_0^\infty d\kappa\,\kappa K_0(0),
\ee
if we use the addition theorem (\ref{addthm}) 
in the singular limit $\rho'\to\rho$.
Such a term, a singular constant, would seem to be a physically
irrelevant contact term, so the two versions of the surface energy
appear to be equivalent.

\begin{acknowledgments}
We are grateful to the US National Science Foundation, grant No.~PHY-0554926,
and the US Department of Energy, grant No.~DE-FG01-03ER03-02,
for partial financial support of this research.  KAM thanks the Physics
Department at Washington University, St.~Louis, for its hospitality during
the early stages of this work.  KK acknowledges support by the Baylor
University Summer Sabbatical Program and by the Baylor University Research
Committee.
We thank Steve Fulling, Prachi Parashar, K.V.~Shajesh, and Jef Wagner
 for helpful conversations, and August Romeo for comments.
\end{acknowledgments}

\end{document}